\documentclass[%
 reprint,
 amsmath,amssymb,
 aps,
]{revtex4-1}

\usepackage{graphicx}
\usepackage{dcolumn}
\usepackage{bm}

\usepackage{upgreek}

\newcommand{\printfnsymbol}[1]{%
  \textsuperscript{\@fnsymbol{#1}}%
}

\begin{document}

\preprint{APS/123-QED}

\title{High coherence superconducting microwave cavities with indium bump bonding}

\author{Chan U Lei,$^*$ Lev Krayzman}
\thanks{These authors contributed equally to this work.}
\email{chanu.lei@yale.edu, lev.krayzman@yale.edu}
\author{Suhas Ganjam}
\author{Luigi Frunzio}
\author{Robert J. Schoelkopf}
\email{robert.schoelkopf@yale.edu}

\affiliation{Department of Applied Physics, Yale University, New Haven, Connecticut 06511, USA}
\affiliation{Yale Quantum Institute, Yale University, New Haven, Connecticut 06520, USA}

\begin{abstract}
Low-loss cavities are important in building high-coherence superconducting quantum computers.  Generating high quality joints between parts is crucial to the realization of a scalable quantum computer using the circuit quantum electrodynamics (cQED) framework. In this paper, we adapt the technique of indium bump bonding to the cQED architecture to realize high quality superconducting microwave joints between chips. We use this technique to fabricate compact superconducting cavities in the multilayer microwave integrated quantum circuits (MMIQC) architecture and achieve single photon quality factor over 300 million or single-photon lifetimes approaching 5 ms. To quantify the performance of the resulting seam, we fabricate microwave stripline resonators in multiple sections connected by different numbers of bonds, resulting in a wide range of seam admittances. The measured quality factors combined with the designed seam admittances allow us to bound the conductance of the seam at $g_\text{seam} \ge 2\times 10^{10} /(\Omega \text{m})$. Such a conductance should enable construction of micromachined superconducting cavities with quality factor of at least a billion. These results demonstrate the capability to construct very high quality microwave structures within the MMIQC architecture.
\end{abstract}

\maketitle

\section{introduction}

Circuit quantum electrodynamics (cQED) is one of the most promising platforms for quantum computation.
Coherence times have been dramatically improved in the past decade \cite{Kjaergaard2019}. Additionally, circuits with dozens of qubits have been realized and used to demonstrate many interesting results, such as molecular simulations \cite{wang2019quantum, kandala2019error, o2016scalable}, condensed matter simulations \cite{ma2019dissipatively}, and proof-of-principle quantum supremacy calculations \cite{arute2019quantum}.
However, further scaling up the number of circuit components while maintaining or even improving their coherence is very challenging.

In the past several years, techniques from the MEMS industry have been applied to the cQED platform to construct multilayer microwave integrated quantum circuits (MMIQC) \cite{Brecht2016MMIQC}. Tremendous progress has been achieved in 3D integration of quantum circuit elements while maintaining their coherence \cite{rosenberg20173d, dunsworth2017low, Brecht2015, brecht2017}. Micromachined superconducting cavities can be highly useful in the MMIQC architecture. Such structures can serve as long-lived quantum memories or as enclosures to suppress radiation loss to the environment and crosstalk between quantum circuit elements. A crucial requirement for constructing high quality micromachined superconducting cavities is the fabrication of high quality microwave joints between layers of the MMIQC \cite{Brecht2015}. The loss associated with joints significantly limits the choice of geometry and materials of the superconducting cavity, as well as the layout of quantum circuits. To maintain the performance of superconducting cavities without sacrificing design flexibility, reducing the loss associated with the joints is critically important.
    
In this work, we adapt indium bump bonding to the cQED architecture. Using this technique, we create superconducting joints with very low loss at microwave frequency. In Sec. 2, we quantify the loss of the resulting joint by fabricating bump-bonded indium stripline resonators and measuring their internal quality factors. In Sec. 3 we apply this indium bump-bonding technique to realize high quality micromachined cavities and show devices with low-power quality factor of over 300 million.

\section{Ultra-low loss joints with indium bump bonding}

Building superconducting cavities requires joining together at least two different parts. 
However, the seam at the joint may limit the coherence of a cavity resonator.
In an ideal joint, the contact surfaces participating in the seam would be identical to the bulk material that comprises the remainder of the cavity; there would be no extra loss associated with the seam.
In a realistic joint, imperfections such as lattice defects, metal oxides, or organic residue on the surfaces can reduce the electrical conductivity across seam and degrade the quality of the resonator.
The loss of the seam can be quantified with the following phenomenological model from \cite{Brecht2015}: 
\begin{equation}
    \frac{1}{Q_\text{seam}} = \frac{1}{g_\text{seam}} \times \frac{\int_\text{seam}|\vec{J_\text{s}} \times \hat{l}|^2 dl}{
    \omega \int_\text{tot}\mu|\vec{H}|^2 dV} \equiv \frac{1}{
    g_\text{seam}} \times  y_\text{seam} ,
\end{equation}
where $\hat{l}$ is a unit vector along the seam, and the integral in the denominator is taken over the entire volume of the mode.
The conductance per unit length $g_\text{seam}$ is an empirical value, and the admittance per unit length $y_\text{seam}$ is a geometric factor that can be calculated analytically or numerically.

\begin{figure*}
\centering
\includegraphics{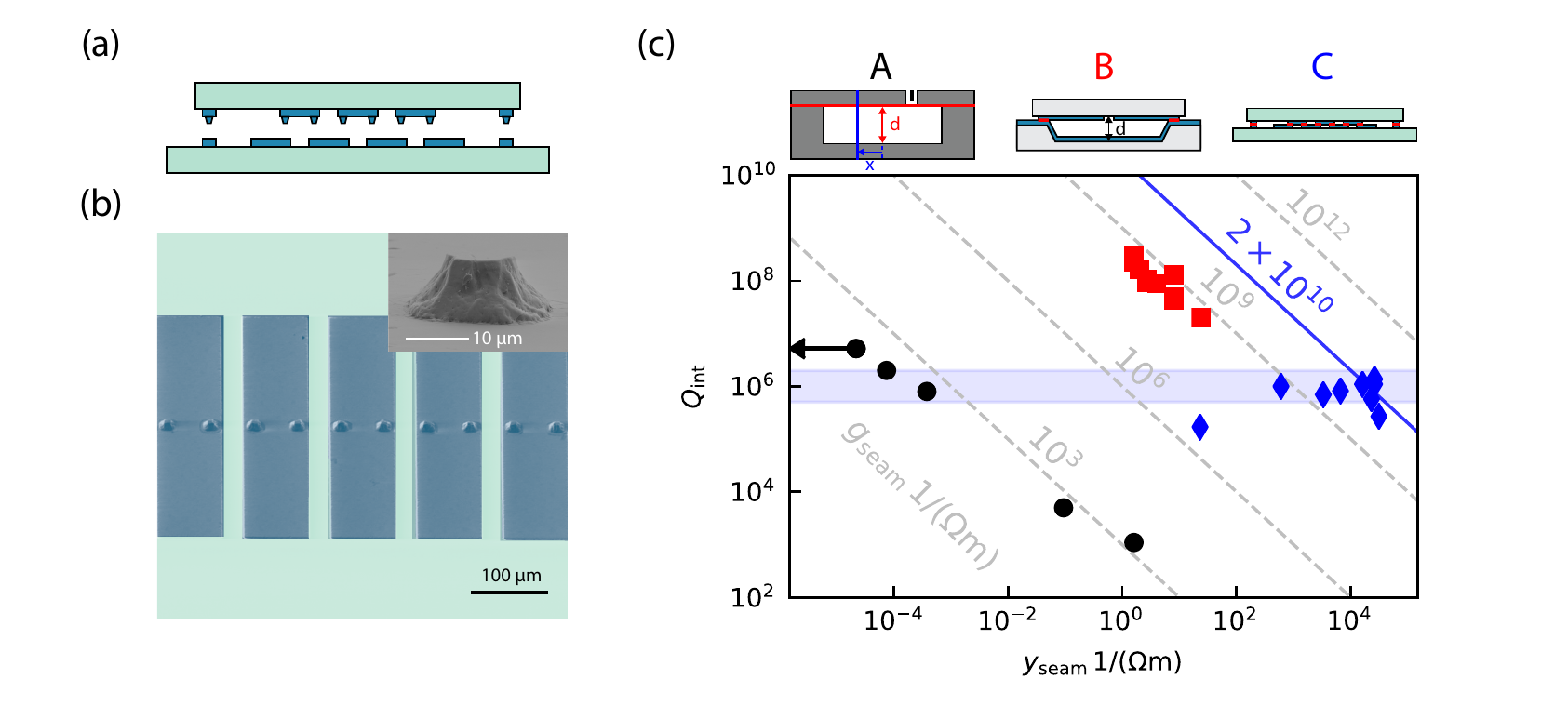}
\caption{(a) Schematic of the bump-bonded stripline resonator. 
The two chips are depicted separately, with sapphire in teal and indium in blue.
The multi-section stripline is visible, with bumps on the daughter chip making the connections.
Additionally, mechanical support bumps can be seen on the edges of the stripline.
(b) False-colored optical micrograph of the daughter chip, with the interrupted stripline together with the bumps visible. 
The inset shows an SEM image of a bump.
(c) Plot of measured internal Q vs simulated $y_\text{seam}$ for three series of devices.
Dashed diagonal lines are lines of constant $g_\text{seam}$.
Series A (black circles) represents TE101 modes of traditionally machined aluminum 6061 cavities with seams placed in varying locations.
The arrow indicates that that device has a $y_\text{seam}$ of nominally 0, although machining imprecision can give it a non-zero $y_\text{seam}$.
Series B (red squares) represents micromachined cavities of differing depths.
Although the points seem to lie along a line of constant $g_\text{seam}$, we note that several other loss mechanisms (metal-air interface, conductor loss) scale with depth in the same way as $y_\text{seam}$, meaning that the cavity could be limited by any or several of these mechanisms.
Series C (blue diamonds) represents the bump-bonded stripline resonators, with varying numbers of bumps.
The blue band is drawn around a control device, with width indicating normal device-to-device variation.
The blue line of $g_\text{seam}=2\times 10^{10}/(\Omega \text{m})$ is a lower bound on the seam conductance that is representative of the several devices with very large seam admittance.
}
\label{fig:fig1}
\end{figure*}

Previous studies \cite{Brecht2015} have shown that seam loss can be dominant in a cavity, depending on cavity geometry and location as well as quality of the seam, as shown in Fig.~\ref{fig:fig1}(c).
Series A (black circles, data taken from \cite{Brecht2015}) represents the TE110 modes of a set of rectangular cavity resonators machined out of bulk 6061 aluminum.
Each device is nominally identical, except for the location of its seam.
The different seam positions result in different values of $y_\text{seam}$, with higher $y_\text{seam}$ corresponding to lower $Q$.
The internal quality factors of these devices lie on a diagonal line corresponding to a $g_\text{seam}\approx 10^3\,/(\Omega\text{m})$.
Since the geometries of the devices are nominally identical, the energy participation in other loss mechanisms remains fixed while the $y_\text{seam}$ varies. 
This is consistent with their Q's being limited by the seam loss.
For micromachined superconducting cavities made by bonding two wafers together, the location of the joint is at the worst location for the fundamental mode, with $y_\text{seam} \approx 1 /(\Omega m)$.
In order to improve its coherence time, it is therefore necessary to increase the $g_\text{seam}$ by developing better joints.

An ideal joint would have both low electrical loss at microwave frequencies and mechanical stability at cryogenic temperatures.
Indium bump bonding is an established method that has been used in the field of cryogenic detectors \cite{denigris2018fabrication}, and has several properties that make it a good candidate for an ultra-low loss joint for superconducting quantum computing applications.
Indium cold welds to itself and is not brittle at cryogenic temperatures \cite{datta2004microelectronic}, allowing for room-temperature bonding which remains robust even with thermal cycling to milli-Kelvin temperatures. 
Additionally, indium superconducts at 3.4~K and can thus form low-loss bonds.
Finally, it is compatible with standard lithographic techniques, enabling the adaptation of existing approaches for scaling up quantum circuits.
Recently, indium bonding has been used for chip hybridization and interconnections \cite{foxen2017qubit, rosenberg20173d, OBrien2017, McRae2017}.
In this work, we optimize the existing bonding technique to attain ultra-low loss joints in the microwave frequency range.

We fabricate indium bumps using a variant of the standard process (see supplement for more details).
Photolithography is used to lift off thermally evaporated indium, forming approximately $15 \, \upmu\text{m} \times 15 \, \upmu\text{m} \times 10 \,\upmu\text{m}$ bumps (see inset in Fig.~\ref{fig:fig1}(b)).
For the other side of the joint, we leave a uniform layer of evaporated indium.
Before bonding, we treat both chips with a plasma surface treatment in order to remove oxide and passivate the surface.
Finally, we bond the two chips in a commercial wafer bonder at room temperature. 

In order to characterize the quality of the resulting joint, we create indium stripline resonators in multiple sections connected by varying numbers of bump bonds (see Fig.~\ref{fig:fig1}(a)).
By varying the number of bumps, we can change the $y_\text{seam}$ of the devices.
The highest $y_\text{seam}$ is achieved when there are the maximum number of bumps given fabrication constraints (see Fig.~\ref{fig:fig1}(b)).
These devices can attain much higher $y_\text{seam}$ than micromachined cavities without too much of a decrease in Q, allowing us to place a tighter lower bound on $g_\text{seam}$.
A control device is also created by placing the entire stripline on one chip, which is still bonded to the second chip by the mechanical support bumps on the edge.
Since such a device has no seam at all, its $y_\text{seam}$ should be precisely 0.

We measure the stripline resonators in a multiplexed package in hanger configuration at the base of a dilution refrigerator, with temperature around 15 mK \cite{axline2016}.
Series C (blue diamonds) in Fig.~\ref{fig:fig1}(c) represents these devices.
A shaded blue band is drawn around the control device with width set by normal variation between devices.
We note that most of the devices lie within this band, except for two with lower Q, which could be caused by some imperfection in the device or the bond.
This indicates that the measured Q factors of the bump-bonded resonators are not limited by the seam loss and are instead limited by other mechanisms, such as dielectric or conductor losses. 
With these data we can place a lower bound on $g_\text{seam} \ge 2\times10^{10}/(\Omega \text{m})$, which is representative of the distribution of several points around our highest $y_\text{seam}$.
A seam with such a conductance would enable a micromachined cavity to have Q of 1-10 billion, depending on depth.

\section{High-quality superconducting micromachined cavities}

\begin{figure*}
  \centering
  \includegraphics{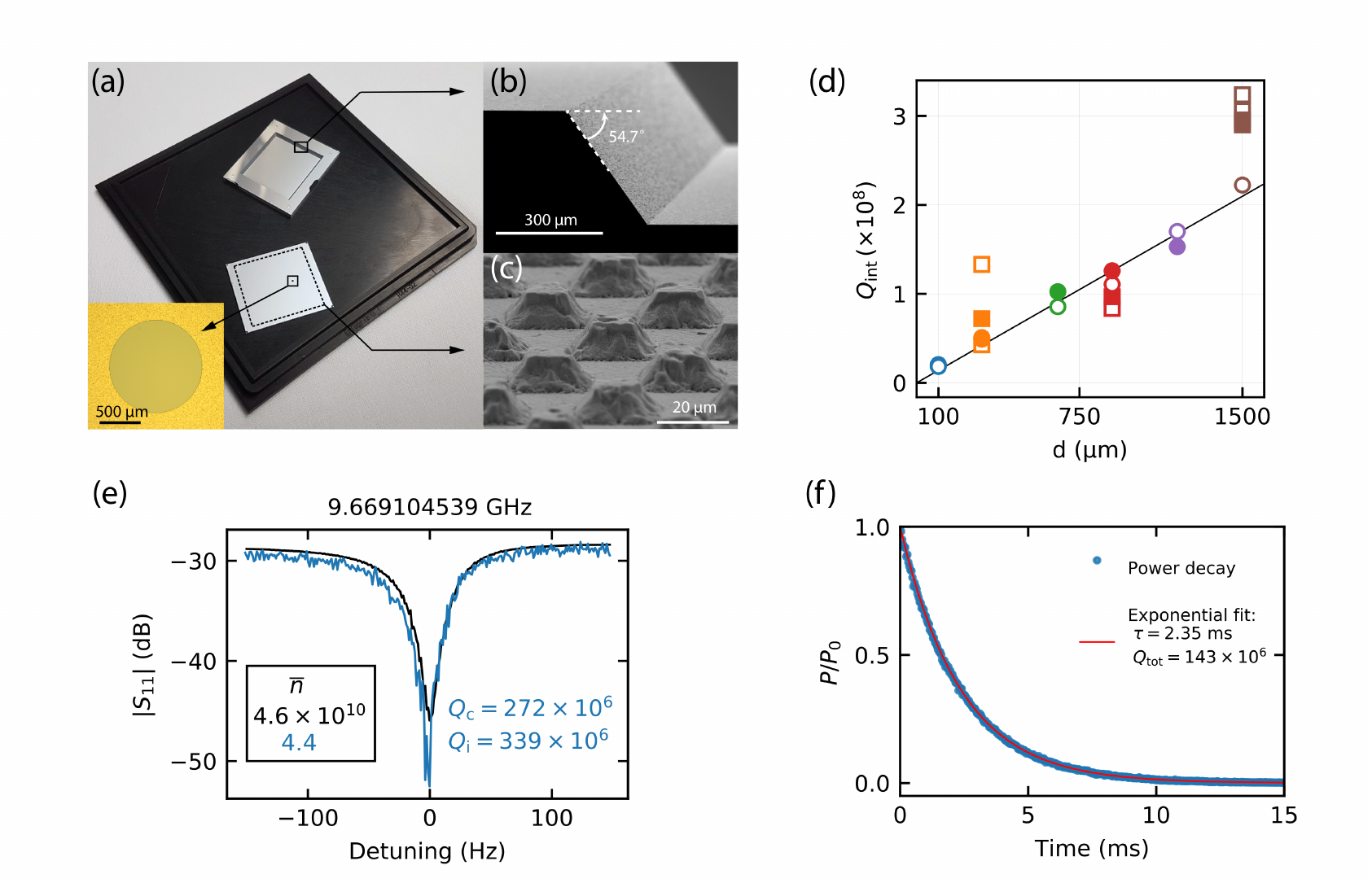}
  \caption{(a) The top chip and the bottom chip of a micromachined cavity before bonding, the dashed line on the top chip indicates the location of the indium bumps, the inset is the optical micrograph of the coupling aperture on the top chip. (b) Side wall of the bottom chip. (c) Electron micrograph of the indium bumps. (d) Internal quality factor of micromachined cavities with cavity depth equal to $100\,\upmu\text{m}$ (blue), $300\,\upmu\text{m}$ (orange), $650\,\upmu\text{m}$ (green), $900\,\upmu\text{m}$ (red), $1200\,\upmu\text{m}$ (purple), and $1500\,\upmu\text{m}$ (brown). Squares (circles) represent devices from batch 1 (2). Opened and filled symbols represent data from two cooldowns that are separated by more than 3 months. Solid line is a linear fit for devices from batch 2 (e) Transmission spectrum of a micromachined cavity with $d = 1500\,\upmu\text{m}$ at $4$ photons (blue) and $10^{10}$ photons (black). (f) Power decay measurements data (blue) and the exponential fit (red).}
  \label{fig:fig2}
\end{figure*}

The high-quality indium joint can be applied to enhance the lifetime of a superconducting cavity resonator.
In this section, we apply the indium bump bonding technique described in the previous section to construct high-quality micromachined cavities. 
The cavities are composed of two silicon chips that are metallized with $1.2\,\upmu\text{m}$ of thermally evaporated indium (Fig.~\ref{fig:fig2}a). 
The bottom chip contains a rectangular recess area made using a potassium hydroxide anisotropic silicon etch, resulting in a $54.7^{\circ}$ sidewall (Fig.~\ref{fig:fig2}b). 
The top chip is patterned with a coupling aperture for measurement (inset in Fig.~\ref{fig:fig2}a). 
In order to improve the seam quality, indium bump arrays are fabricated along the contact region on the top chip (dashed line on the top chip in Fig.~\ref{fig:fig2}a). 
The two chips are bonded together with a commercial wafer bonder with a force of $2\,\text{kN}$ in ambient conditions to form a micromachined cavity, which is mounted on a sample holder to be measured in a reflection configuration (see supplementary materials for more details). 

The internal loss of the micromachined cavity consists of the conductive loss of the indium, the dielectric loss of the indium oxide, and the loss from the seam formed at the contact of the two chips
\begin{equation}
    \frac{1}{Q_{\text{int}}} = 
    \frac{\alpha R_s}{\omega\mu_0\lambda_0}+
    p_{\text{diel}}\tan{\delta}+
    \frac{y_{\text{seam}}}{g_{\text{seam}}}.
\end{equation}
Here, $\omega$ is the frequency of the cavity resonance, $\mu_0$ is the vacuum permeability, and $\lambda_0$ is the penetration depth of indium. The kinetic inductance fraction $\alpha$, the surface dielectric participation ratio $p_{\text{diel}}$, and the seam admittance per unit length $y_{\text{seam}}$ are geometrical parameters that can be calculated analytically or numerically. The surface resistance $R_s$ of the superconductor, the loss tangent $\tan{\delta}$ of the surface dielectric, and the seam conductance per unit length $g_{\text{seam}}$ represent the intrinsic loss of each respective loss mechanism.

For the TE110 mode of the micromachined cavity, we observe that the internal quality factor increases linearly with $d$, as shown in Fig.~\ref{fig:fig2}(d).  However, for this mode $\alpha$, $p_{\text{diel}}$, and $y_{\text{seam}}$ are all inversely proportional to the depth of the cavity $d$. Therefore, we cannot determine which loss mechanisms are dominant based on this set of measurements. Nevertheless, by attributing the total loss to a single loss mechanism, we can place bounds on the corresponding intrinsic loss, which gives $R_{\text{s}} \leq 261 \,\text{n}\Omega$, $\tan{\delta} \leq 1.2\times10^{-2}$, and $g_{\text{seam}} \geq 3.4\times10^8\,/(\Omega\text{m})$. The highest internal quality factor we have achieved in this study is $3.4\times10^8$ with cavity depth $d=1.5\,\text{mm}$.
The lifetimes of these cavities are independent of whether a phase-sensitive heterodyne measurement (Fig.~\ref{fig:fig2}(e)) or a phase-insensitive power ringdown measurement (Fig.~\ref{fig:fig2}(f)) is used, i.e., $T_2^{*} \simeq 2 T_1$. No apparent degradation in internal quality factor has been observed after the samples have been exposed to air for over 3 months. 

\begin{figure*}
  \centering
  \includegraphics{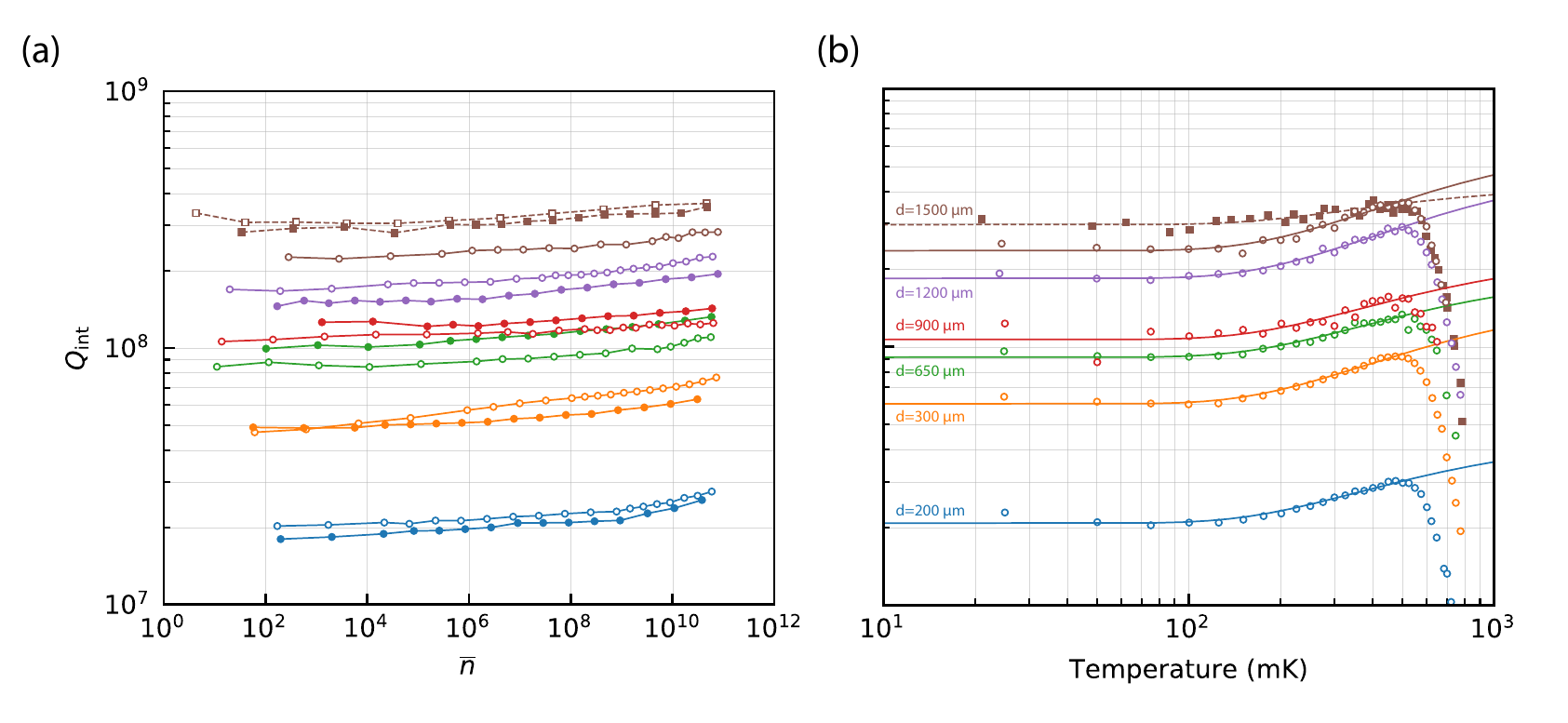}
  \caption{(a) Power dependence of the internal quality factors of micromachined cavities. (b) Temperature dependence of the internal quality factors of micromachined cavities. The solid curves are the fit with TLS model (E.q. (\ref{tandeltaTLS})) for data below $500\,\text{mK}$. Both figures share the same symbol and color codes used in  Fig.~\ref{fig:fig2}(d)}
  \label{fig:fig3}
\end{figure*}

As in other 3D superconducting cavity resonators, the internal quality factors $(Q_{\text{int}})$ of the micromachined cavities depend very weakly on the average photon number $(\overline{n})$.
We observed an increase of less than $60\%$ from several to 10 billion photons (see Fig.~\ref{fig:fig3}(a)). 
This represents a marked difference from coplanar waveguide (CPW) resonators, the internal quality factors of which typically increase by more than a factor of two from one to $10^5$ photons \cite{pappas2011TLS, quintana2014Al, bruno2015NbTiN, calusine2018TiN}, at which point the increase becomes saturated by the power-independent losses. 
This power-dependent behavior of the internal quality factor of superconducting microwave resonators is generally believed to be caused by the saturation of the two-level system (TLS) dissipators by the electric field of the resonator. These TLS's are believed to be hosted by the surface dielectrics of the resonator \cite{gao2008thesis, pappas2011TLS}. The resulting power- and temperature-dependent surface dielectric loss tangent is given by
\begin{equation}
    \tan{\delta_{\text{TLS}}} = \tan{\delta^0_{\text{TLS}}} \frac{ \tanh{\left(\frac{\hbar\omega}{2k_{\text{B}}T}\right)}}{\sqrt{1+\frac{\overline{n}}{n_\text{c}}}}, \label{tandeltaTLS}
\end{equation}
where $k_{\text{B}}$ is the Boltzmann constant and $\tan{\delta_{\text{TLS}}^{0}}$ is the loss tangent of the TLS dissipators at zero temperature and zero electric field. 
The critical photon number $n_\text{c}$ is the minimum number of photons required to saturate the TLS loss. For typical aluminum CPW resonators, $n_{\text{c}}\simeq 500$ \cite{quintana2014Al, dunsworth2017Al}, which is determined by the properties of the TLS's in the hosted material as well as the electric field-weighted mode volume of the resonance $V_{\text{eff}}^{\text{E}}$. Assuming that the dipole moment and the coherence time of the TLS's remain unchanged, $n_{\text{c}} \propto  V_{\text{eff}}^{\text{E}}/\omega$.
We note that the mode volume of a micromachined cavity is several million times larger than that of a CPW resonator. As shown in Fig.~\ref{fig:fig3}(a), $Q_\text{int}$ of the micromachined cavities starts to increase around $10^9$ photons, i.e., $n_{\text{c}}\simeq 10^9$. Therefore, the ratio $n_{\text{c}}/V_{\text{eff}}^{\text{E}}$ is  about the same for CPW resonators and micromachined cavities, suggesting that the TLS properties are similar for aluminum and indium surfaces.

In addition, the TLS dissipators can be saturated by thermal excitations, as can be seen in Eq.(\ref{tandeltaTLS}).
As the temperature increases, the $Q_{\text{int}}$ of the micromachined cavities rises and is maximized at $T\simeq 500\,\text{mK}$, then drops due to the increase of broken Cooper pairs, (see Fig.~\ref{fig:fig3}(b)).  This temperature-dependent behavior indicates that TLS loss is one of the dominant loss channels in these cavities. 
Assuming the TLS dissipators are homogeneously distributed within a thin dielectric layer on the surface of the micromachined cavities, the internal loss of the cavities can be expressed as the sum of the temperature-independent loss and the temperature-dependent TLS loss via
\begin{equation}
    \frac{1}{Q_{\text{int}}} = \frac{1}{Q_{\text{int},0}}+ p_{\text{diel}}\tan{\delta^0_{\text{TLS}}}\tanh{\left(\frac{\hbar\omega}{2k_{\text{B}}T}\right)}, \label{QintTLS}
\end{equation}
here we consider $\overline{n} \ll n_\text{c}$ and $T \ll T_{\text{c}}$, with the superconducting transition temperature $T_{\text{c}}=3.4\,\text{K}$ for indium.
The solid curves in Fig.~\ref{fig:fig3}(b) are the fit with E.q. (\ref{QintTLS}) assuming a $3\,\text{nm}$-thick dielectric layer with $\epsilon_r=10$, which gives $\tan{\delta^0_{\text{TLS}}} = \left(5.6\pm 1.6\right)\times10^{-3}$ on the indium surface. This is similar to the loss tangent of the TLS dissipators found on aluminum oxide, which have $\tan{\delta^0_{\text{TLS}}} \simeq 1.0\times10^{-3}$ \cite{pappas2011TLS, mcrae2019dielectric}.

\section{Conclusion and perspectives}

In this work, we have demonstrated techniques to fabricate and characterize high-quality superconducting joints. 
We have used the bump-bonding technique to construct high quality indium micromachined cavities using industry standard techniques that do not require manual assembly and is fully compatible with existing Josephson junction fabrication processes. 
We have achieved a low-power internal quality factor of over 300 million corresponding to an intrinsic T1 approaching 5 ms, which is comparable to the performance of conventionally machined cavities used for quantum memories in cQED. 
This result together with the ability to make low-loss joints shows that indium is a good superconductor for making high-quality quantum circuits. 
Integration of a micromachined cavity with a qubit has been demonstrated \cite{brecht2017}; by applying the techniques developed here we can improve the lifetime of a MMIQC-based quantum memory to the millisecond level.

The desire to build 3D cavities either to serve as quantum memories or to provide electromagnetic shielding highlights the fact that seam losses are very important for the 3D integration of quantum circuits. Since the scaling of superconducting circuits for quantum information processing will likely require multilayer circuits and interconnects \cite{brecht2017, rosenberg20173d}, the improved performance of the superconducting joints demonstrated here can be a key enabler for enhancing coherence in a wide range of devices. It also allows the realization of low-loss compact multilayer circuits including flip-chip resonators and lumped-element microwave networks.
This work thus provides an important step towards building a more complicated MMIQC while improving its performance.
\newline

\section{Acknowledgements}
We thank Harvey Moseley, Ari Brown, Nicholas Costen, Timothy Miller, and Luke Burkhart, and Gianluigi Catelani for useful conversations; Jan Schroers and Teresa Brecht for assistance with wafer bonding; Charles Ahn, Frederick Walker, and Cristina Visani for assistance with DC measurements of our indium films; Philip Reinhold and Christopher Axline for experimental assistance; Michael Power, Michael Rooks, Christopher Tillinghast, James Agresta, Yong Sun, Sean Rinehart, and Kelley Woods for assistance with device fabrication.
This research was supported by the U.S. Army Research Office grant W911NF-18-1-0212. S. G. was supported by a Max Planck Research Award from the Alexander von Humboldt Foundation.
Facilities use was supported by the Yale university cleanroom and YINQE.

\bibliography{reference} 

\end{document}


\preprint{APS/123-QED}

\title{Supplementary Materials: High coherence superconducting microwave cavities with indium bump bonding}

\author{Chan U Lei,$^*$ Lev Krayzman}
\thanks{These authors contributed equally to this work.}
\email{chanu.lei@yale.edu, lev.krayzman@yale.edu}
\author{Suhas Ganjam}
\author{Luigi Frunzio}
\author{Robert J. Schoelkopf}
\email{robert.schoelkopf@yale.edu}

\affiliation{Department of Applied Physics, Yale University, New Haven, Connecticut 06511, USA}
\affiliation{and Yale Quantum Institute, Yale University, New Haven, Connecticut 06520, USA}

\maketitle
\beginsup

\section{Fabrication and measurement of micromachined cavities}

\begin{figure*}
  \centering
  \includegraphics{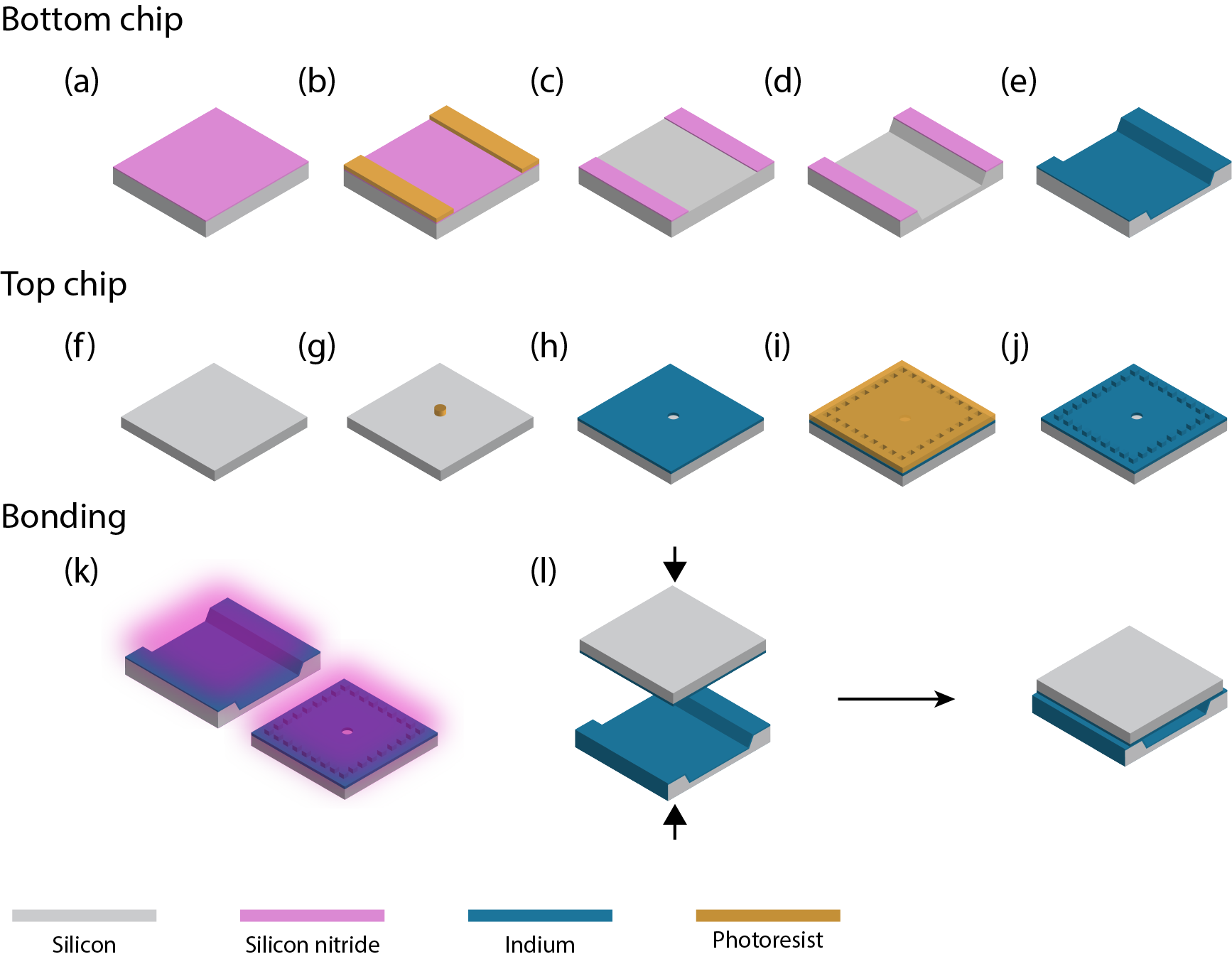}
  \caption{Fabrication process of the micromachined cavity.
  (a) A layer of $300\,\text{nm}$-thick silicon nitride is grown with PECVD on the (100) surface of a silicon substrate for the bottom chip. (b) The silicon nitride layer is photolithographically patterned. (c) The pattern is transferred to the silicon nitride by $\text{CHF}_3/\text{O}_2$ RIE. The photoresist is removed after the etch. (d) Using the patterned silicon nitride as a mask, a recess area is created in the silicon chip using an anisotropic KOH etch. (e) After removing the silicon nitride with $\text{CHF}_3/\text{O}_2$ RIE, $1.2\,\mu\text{m}$ of indium is thermally evaporated on the bottom chip.
  (f) The silicon wafer for the top chip fabrication is solvent cleaned. (g) The coupling aperture is photolithographically patterned. (h) A $1.2\,\mu\text{m}$ film of indium is deposited and lifted off. (i) The array of indium bumps is photolithographically patterned. (j) The indium oxide on the exposed indium layer is removed with argon ion milling, then $8\,\mu\text{m}$ of indium is deposited and lifted off to form the indium bumps. 
  (k) A downstream atmospheric plasma process of mixed helium, nitrogen, and hydrogen is applied to the surface of of both chips in order to remove the oxide and passivate the surfaces. (l) The two chips are bonded with a flip-chip wafer bonder at $2\,\text{kN}$ (effective pressure of several hundred MPa) to form the micromachined cavity.}
  \label{fig:figS1}
\end{figure*}

\begin{figure*}
  \centering
  \includegraphics{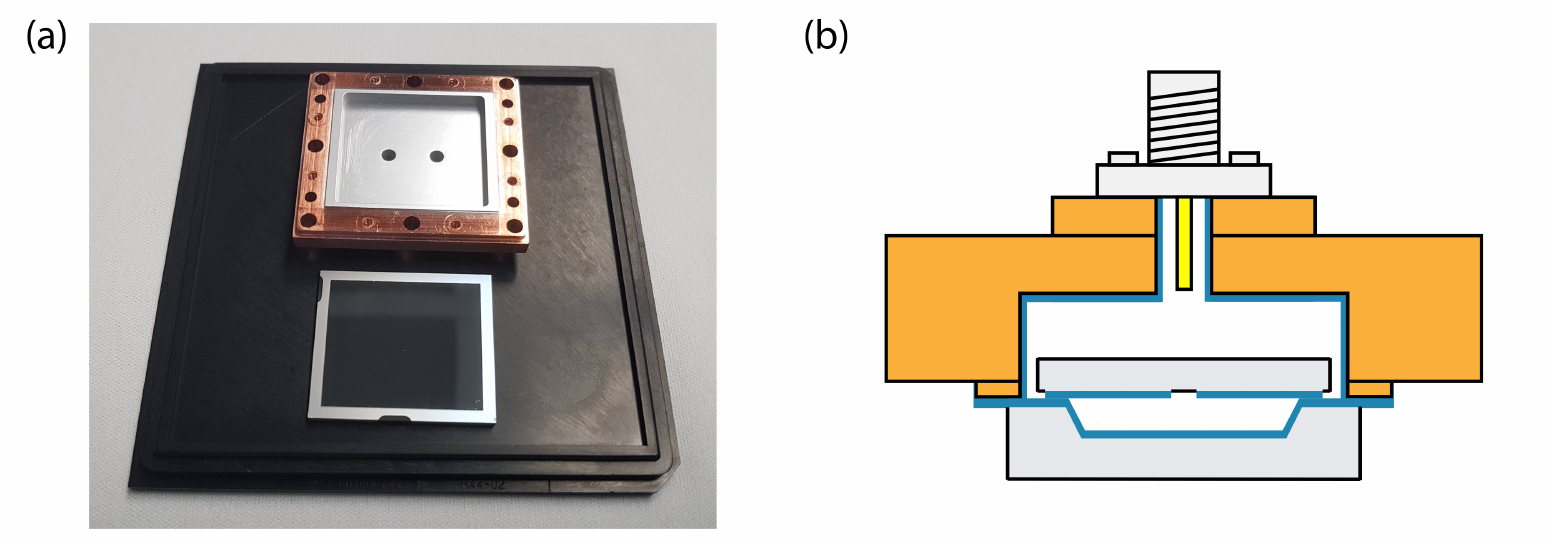}
  \caption{(a) photograph of a sample holder and a micromachined cavity before mounting. (b) A schematic diagram of the cross section of a micromachined cavity mounted on a sample holder. The sample holder is made out of OFHC copper, whose inner surface is covered with $10\,\upmu\text{m}$ of indium. The coupling to the resonances in the micromachined cavity can be controlled by the length of the pin connected to the SMA connector.}
  \label{fig:figS2}
\end{figure*}

The fabrication process of the micromachined cavity is a modification of the process described in the supplement of \cite{Brecht2015} and is shown in Fig.~\ref{fig:figS1}. This consists of three main steps: fabrication of the bottom chip, fabrication of the top chip, and flip-chip bonding. The fabrication of the bottom chip is depicted on Fig.~\ref{fig:figS1}(a) through Fig.~\ref{fig:figS1}(e). A layer of $300\,\text{nm}$-thick silicon nitride is first deposited on the (100) plane of a silicon substrate with plasma enhanced chemical vapor deposition (PECVD) (Fig.~\ref{fig:figS1}(a)). Then, the silicon nitride layer is patterned by etching with $\text{CHF}_3/\text{O}_2$ reactive ion etch (RIE) through a photolithographic mask (Fig.~\ref{fig:figS1}(b) and (c)). A cavity is formed into the silicon wafer by a potassium hydroxide (KOH) anisotropic etch with the patterned silicon nitride layer as a mask (Fig.~\ref{fig:figS1}(d)). After the KOH etch, the silicon nitride mask is removed and a layer of $1.2\,\upmu\text{m}$ of indium is thermally evaporated on the bottom chip (Fig.~\ref{fig:figS1}(e)).

The fabrication process of the top chip is depicted on Fig.~\ref{fig:figS1}(f) through Fig.~\ref{fig:figS1}(j). First, a  coupling aperture is patterned on a solvent-cleaned silicon substrate using photolithography (Fig.~\ref{fig:figS1}(f) and (g)). Then, a layer of $1.2\,\upmu\text{m}$-thick indium is deposited and lifted off to form the top of the micromachined cavity (Fig.~\ref{fig:figS1}(h)). In order to improve the quality of the joint between the two parts of the cavity, an array of indium bumps is photolithographically patterned on top of the existing indium layer (Fig.~\ref{fig:figS1}(i)). Argon ion milling is applied in high vacuum to remove the indium oxide on the exposed indium surface. Then, $8\,\upmu\text{m}$ of indium is deposited without breaking vacuum and subsequently lifted off to form the indium bumps (Fig.~\ref{fig:figS1}(j)). 

Finally, the flip-chip bonding process is depicted in Fig.~\ref{fig:figS1}(k) and (l). The indium surfaces of the top and bottom chips are treated with an atmospheric plasma surface treatment (OES Ontos7) comprising mixed helium, nitrogen, and hydrogen immediately prior to bonding (Fig.~\ref{fig:figS1}(k)). The two chips are then bonded together at ambient conditions at $2\,\text{kN}$ with a dedicated flip-chip wafer bonder (Fig.~\ref{fig:figS1}(l)).

After fabrication, the micromachined cavity is mounted to a sample holder for measurement in reflection. As shown in Fig.~\ref{fig:figS2}, the sample holder contains an SMA connector with a pin which couples to the micromachined cavity through its coupling aperture. The length of the coupling pin is designed to attain critical coupling to the TE101 mode of the micromachined cavity in order to obtain a maximum signal-to-noise ratio. The sample is mounted on the base stage of a cryogen-free dilution refrigerator and shielded with Cryoperm magnetic shielding. The input microwave signal passes through $20\,\text{dB}$ and $30\,\text{dB}$ attenuators on the $4\,\text{K}$ stage and the base stage of the dilution refrigerator, respectively, in order to reduce thermal noise. The signal then enters the $-10\,\text{dB}$ coupling port of a directional coupler and is directed to the input port of the sample. The input microwave signal is reflected from the sample, $90\%$ of the reflected signal is transmitted through the transmitted port of the directional coupler, then passes through two isolators at $20\,\text{mK}$ and is amplified by a cryogenic HEMT amplifier at $4\,\text{K}$. The resulting output signal is further amplified with a room-temperature low noise amplifier for analysis.

\section{Energy loss in micromachined cavities}
The reciprocal of the internal quality factor is defined as the fraction of the total energy lost per cycle. The total loss can be expressed as the sum over the different loss channels via
\begin{equation}
    \frac{1}{Q_{\text{int}}} = \frac{P_{\text{diss}}}{\omega U_{\text{tot}}} = \sum_{i}{\frac{P_{\text{diss}, i}}{\omega U_{\text{tot}}}} = \sum_{i}{\frac{p_i}{q_i}},
\end{equation}
where $p_{i}$ is the energy participation ratio and $q_i$\textendash{}the quality factor of the loss channel $i$. The energy participation ratio is the fraction of the total energy stored in the loss element $i$. This depends on the geometry of loss channel $i$ and the mode shape of the resonator. The quality factor is determined by the loss properties of the material of $i$. For the micromachined cavity, the losses are dominated by three loss channels: conductive loss in the superconductor, dielectric loss in the very thin layer of dielectric material on cavity surface, and loss from the current flow across the seam formed by the joint between the two parts of the cavity. 

In the following, we will focus our discussion on the participation ratios and the material quality factors associated with these three type of losses for the TE101 mode of a rectangular cavity. The electromagnetic fields of the TE101 resonance of a rectangular cavity (with lateral dimensions $a$, $b$, and thickness $d$) are
\begin{align}
    E_{z} &= E_0 \sin{\left(\frac{\pi x}{a}\right)} \sin{\left( \frac{\pi z}{b}\right)}, \\
    H_{x} &= -i\frac{\pi E_0}{k \eta b} \sin{\left(\frac{\pi x}{a}\right)} \cos{\left(\frac{\pi z}{b}\right)}, \\
    H_{y} &= i\frac{\pi E_0}{k \eta a} \cos{\left(\frac{\pi x}{a}\right)} \sin{\left(\frac{\pi z}{b}\right)},
\end{align}
where $\eta=\sqrt{\frac{\mu_0}{\epsilon_0}}$, $k=\frac{\omega}{c}=\sqrt{\left(\frac{\pi}{a}\right)^2+\left(\frac{\pi}{b}\right)^2}$. Given the electromagnetic fields of the cavity resonance, the participation factors of the loss channels can be calculated. 

For a micromachined cavity, the lateral dimensions $a$ and $b$ are much larger than the thickness $d$, giving the participation factor of the conductive loss as
\begin{equation}
    p_{\text{cond}} = 
    \frac{\lambda \int_\text{surf}\left|\vec{H}_{\parallel}\right|^2d\sigma}
    {\int_\text{vol}\left|\vec{H}\right|^2dv} = 
    \frac{\omega \mu_0 \lambda}{\eta} \frac{2\pi^2}{k^3} \left(\frac{1}{b^2}+\frac{1}{a^2}\right)\frac{1}{d},
    \label{eq:pcond}
\end{equation}
where $\lambda$ is the penetration depth of the superconductor. The corresponding material quality factor $q_{\text{cond}}=X_s/R_s=\omega\mu_0\lambda/R_s$ is the ratio of the surface reactance to the surface resistance of the superconductor. 

The participation factor of the surface dielectric loss is
\begin{equation}
    p_{\text{surf}} = 
    \frac{t_{\text{ox}} \int_{\text{surf}}\epsilon\left|\vec{E}\right|^2d\sigma}
    {\int_\text{vol}\epsilon\left|\vec{E}\right|^2dv} = \frac{2t_{\text{ox}}}{\epsilon_r}\frac{1}{d},
\end{equation}
where $t_{\text{ox}}$ is the thickness of the surface dielectric material, which is assumed to be $3\,\text{nm}$ in this work. The corresponding material quality factor $q_{\text{surf}}=1/\tan{\delta}$ is the reciprocal of the loss tangent of the surface dielectric material.

For the seam loss, the participation of the loss channel is quantified by the seam admittance per unit length, which is 
\begin{equation}
    y_{\text{seam}} = 
    \frac{\int_{\text{seam}}\left|\vec{J}_{\text{surf},\perp}\right|^2 dl}
    {\omega\mu_0\int_\text{vol}\left|\vec{H}\right|^2 dv} = 
    \frac{4\pi^2}{\eta k^3}\left(\frac{1}{b^3}+\frac{1}{a^3}\right)\frac{1}{d},
\end{equation}
and the quality of the seam is quantified by the seam conductance per unit length $g_{\text{seam}}$.

Since the conductive loss, the surface dielectric loss, and the seam loss are all proportional to the reciprocal of the cavity thickness, the linear fit in Fig.~2(e) in the main text can only permit the extraction of the upper bounds for $R_s$ and $\tan{\delta}$ and the lower bound for $g_{\text{seam}}$, which gives $R_{\text{s}} \leq 261 \,\text{n}\Omega$, $\tan{\delta} \leq 1.2\times10^{-2}$, and $g_{\text{seam}} \geq 3.4\times10^8\,1/(\Omega\text{m})$.
\newline

\section{Extracting critical temperature and penetration depth with Mattis-Bardeen theory}

We extract the critical temperature and penetration depth of the indium film forming the micromachined cavity by fitting the frequency of the resonant mode to the Mattis-Bardeen model \cite{mattis1958theory}.
We follow the method of \cite{reagor2013reaching}, described in more detail in section 5.1.5 of \cite{reagor2016superconducting}; see also \cite{brecht2017micromachined,tinkham2004introduction, zmuidzinas2012superconducting, gao2008physics}.

First, we numerically calculate the superconducting gap as a function of temperature (in units of $T_c$) for indium, noting that this is not strongly dependent on the Debye temperature of the material.
This is given by the equation
\begin{equation*}
 \frac{1}{N(0)V} = \int_0^1{\frac{\tanh{\left(\frac{1}{2\tau}\sqrt{\xi^2(2\kappa)^2+\delta^2r^2}\right)}}{\sqrt{\xi^2+\delta^2r^2/(2\kappa)^2}}\,\mathrm{d}\xi},
\end{equation*}
where $N(0)$ is the number of Cooper pairs at zero temperature, $V$ is the BCS pairing potential, $\tau$ and $\delta$ are the temperature scaled to $T_c$ and gap scaled to its zero-temperature value $\Delta(0)$, respectively, $\kappa=\frac{\Theta_D}{2T_c}$, and $r=\frac{\pi}{e^{\gamma}}=\frac{\Delta(0)}{k_B T_c}\approx1.76$.
The low-temperature limit gives $\frac{1}{N(0)V}=\log{\left(\frac{4\kappa}{r}\right)}$, assuming $\kappa \gg r$, which allows the above to be solved numerically. 
We then use the result to compute a table of the complex conductivity $\sigma_1(\omega,T) - i \sigma_2(\omega,T)$ as given by Mattis-Bardeen.
From this, we can calculate the surface impedance $Z_S$ of the film.
The surface resistance $R_S=\Re{(Z_S)}$ changes the Q of the mode and the surface reactance $X_S=\Im{(Z_S)}$ shifts the resonant frequency via
\begin{equation}
    Z_s =X_s + i R_s = \frac{\omega \mu_0 \lambda}{p_\mathrm{cond}} \left(\frac{1}{Q_\mathrm{cond}} + 2i\frac{\delta f}{f}\right),
\end{equation}
where $Q_\mathrm{cond}$ is the quality factor due only to conductor losses.
We can fit the frequency shift and the internal $Q$ as functions of temperature to the model with two free parameters that scale the curve in the x and y directions.
The x scaling gives a value of $T_c$, while the y scaling allows us to extract $p_\text{cond}$.
From $p_\text{cond}$, we can obtain $\lambda$ by calculating the integrals in Eq.~\ref{eq:pcond}.
The $Q$ fit has an additional parameter, $Q_0$, that represents the temperature independent part of the total loss.

\begin{figure*}
    \centering
    \begin{subfigure}{ 
        \includegraphics[width=0.47\linewidth]{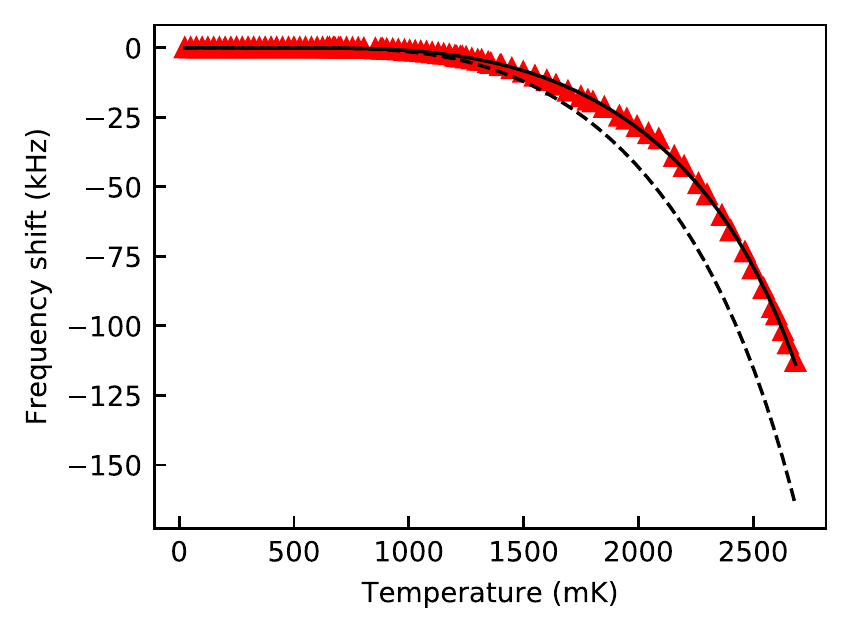}}
    \end{subfigure}
    \begin{subfigure}{
        \includegraphics[width=0.47\linewidth]{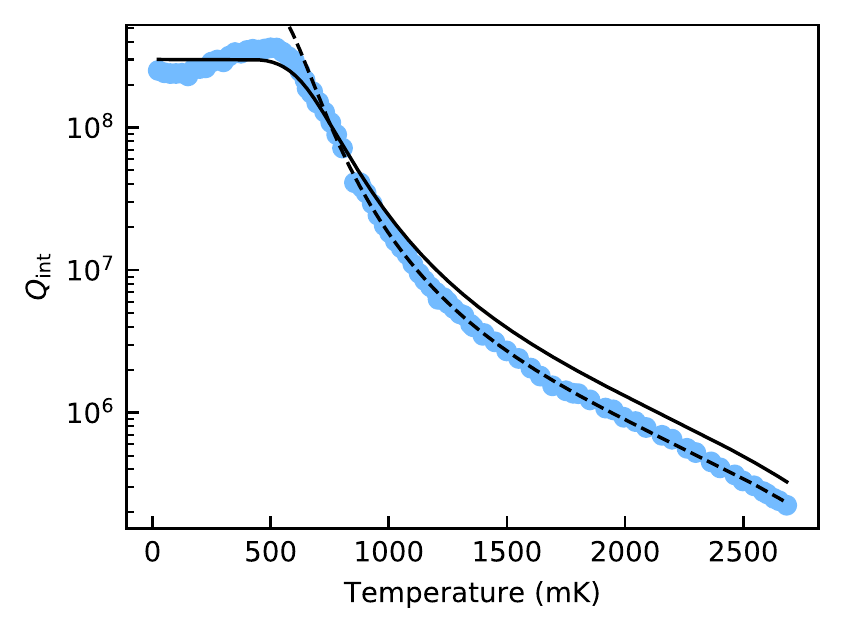}}
\end{subfigure}
    
\centering
  \caption{Plot of fits of the resonant frequency (left) and internal quality factor (right) as a function of temperature of a micromachined cavity of depth 1490 $\upmu\text{m}$. 
  The solid line uses parameters extracted from the frequency fit and the dashed line uses parameters extracted from the $Q$ fit.
  The fit is obtained using the  Mattis-Bardeen result. 
  Note that for the quality factor fit, the lower-temperature values are explained by a temperature-independent $Q_\text{int,0}$ and the temperature-dependent TLS losses (see Fig. 3(b) in the main text), we thus fit $Q_\textrm{int}$ for $T>700\,\textrm{mK}$.
  The frequency fit has two free parameters: $T_c$ and $p_\text{cond}$, which correspond to scaling in the x and y directions, respectively.
  The $Q$ fit has $Q_0$ as an additional parameter.
  The fits presented here assume the local limit of the superconductor; we do not present plots of the fits in the extreme anomalous limit as they would be indistinguishable from these.
  We extract $T_c$ between 3.2 and 3.4 K and $\lambda$ between 32 and 68 nm.
  }
  \label{fig:figS3}
\end{figure*}

The frequency and $Q$ fits of the temperature for our deepest cavity ($d=1490\,\upmu\mathrm{m}$) are plotted in Fig.~\ref{fig:figS3} (fitted separately). 
At higher temperatures, the resonator was too undercoupled to measure and we were not able to obtain data near $T_c$.
As a result, $T_c$ and $p_\text{cond}$ are not very well constrained by the fit.
While our model fits frequency (solid line in Fig.~\ref{fig:figS3}) and $Q$ (dashed line in Fig.~\ref{fig:figS3}) well individually, the two fits result in inconsistent $T_c$ and $p_\text{cond}$. 
In addition, we cannot determine whether our data fits better to a model in the local or dirty limit or to a model in the extreme anomalous or clean limit of the superconductor.
Possible reasons include temperature dependence in other parts of the system, e.g. the seam, or changing transition between anomalous or clean and local or dirty limits of the superconductor with temperature, which is not captured by the model.
We thus report results from both frequency and $Q_\text{int}$ fits, in both local and extreme anomalous limits (see Table~\ref{tab:tableS1}).
We obtain a penetration depth $\lambda$ between 32\textendash{}68 nm and a $T_c$ between 3.17\textendash{}3.35 K.
The penetration depth is comparable to that of aluminum \cite{reagor2013reaching} and the critical temperature is close to the known value of $3.4\,\text{K}$, indicating the suitability of our indium for high quality superconducting circuits.

\begin{table}
\begin{ruledtabular}
    \caption{$T_c$ and $\lambda$ extracted from both frequency and $Q$ fits in both the local and extreme anomalous regimes. 
    Errors reported are the fit errors.}
    \label{tab:tableS1}
    \begin{tabular}{l c c}
         Fit, limit & $T_c\,(\text{K})$ & $\lambda\,(\text{nm})$ \\
         \hline
         Frequency, local & $3.29 \pm 0.01$ & $32.35 \pm 0.66$ \\
         Frequency, anomalous &  $3.17 \pm 0.01$ & $42.98 \pm 0.72$ \\
         $Q$, local &  $3.35 \pm 0.03$ & $51.80 \pm 2.44$ \\ 
         $Q$, anomalous & $3.24 \pm 0.04$ & $67.92 \pm 4.01$ \\
    \end{tabular}
\end{ruledtabular}
\end{table}

\bibliography{apssamp}